\documentclass[oneside]{article}
\usepackage[T1]{fontenc}
\usepackage{authblk}
\usepackage{float}
\usepackage{amsmath}
\usepackage{graphicx}
\usepackage{xfrac}
\usepackage[english]{babel}
\usepackage{lmodern}
\usepackage[T1]{fontenc}
\usepackage[latin9]{inputenc}

\usepackage{csquotes}
\mathchardef\mhyphen="2D

\usepackage{mathtools}
\usepackage{graphicx}
\usepackage{esint}
\usepackage[unicode=true,
 bookmarks=false,
breaklinks=false,pdfborder={0 0 1},backref=false,colorlinks=false]
 {hyperref}
\usepackage[a4paper, total={210mm,297mm},margin=2.0cm]{geometry}

\title{Enhanced computational performance of the lattice Boltzmann model
for simulating micron- and submicron-size particle flows and
non-Newtonian fluid flows}

\author[1]{Hakan Ba\c sa\u gao\u glu \thanks{Electronic address: \texttt{hbasagaoglu@swri.org}; Corresponding author}}
\author[2]{John R. Harwell}
\author[3]{Hoa Nguyen}
\author[4,5,6]{Sauro Succi}

\affil[1]{Mechanical Engineering Division, Southwest Research Institute, San Antonio, TX 78238 USA}
\affil[2]{Defense Intelligence Solutions Division, Southwest Research Institute, San Antonio, TX 78238 USA}
\affil[3]{Department of Mathematics, Trinity University, San Antonio, TX 78212 USA}
\affil[4]{Center for Life Nano Science@La Sapienza, Istituto Italiano di Tecnologia, 00161 Roma, Italy}
\affil[5]{Istituto per le Applicazioni del Calcolo CNR, via dei Taurini 19, Rome, Italy}
\affil[6]{Institute for Applied Computational Science, Harvard John A. Paulson School of Engineering And Applied Sciences, Cambridge, MA 02138, United States}

\date{\today}

\begin{document}

\maketitle

\begin{abstract}
Significant improvements in the computational performance of the lattice-Boltzmann (LB) model, coded in FORTRAN90, were achieved through application of enhancement techniques. Applied techniques include optimization of array memory layouts, data structure simplification, random number generation outside the simulation thread(s), code parallelization via OpenMP, and intra- and inter-timestep task pipelining. Effectiveness of these optimization techniques was measured on three benchmark problems: (i) transient flow of multiple particles in a Newtonian fluid in a heterogeneous fractured porous domain, (ii) thermal fluctuation of the fluid at the sub-micron scale and the resultant Brownian motion of a particle, and (iii) non-Newtonian fluid flow in a smooth-walled channel. Application of the aforementioned optimization techniques resulted in an average 21$\times$ performance improvement, which could significantly enhance practical uses of the LB models in diverse applications, focusing on the fate and transport of nano-size or micron-size particles in non-Newtonian fluids.      
\end{abstract}

\section{Introduction}
\label{intro}

The lattice-Boltzmann (LB) method has emerged as a versatile computational fluid dynamics tool \cite{S01,W00} to simulate mesoscale simulations of flow and transport of nano- and micron-size particles \cite{L94a,DA03,BMSY12, BAS13}, multiphase and multicomponent flows \cite{SC93,AL11,BS13, KGA13}, laminar and turbulent flows \cite{CKO03,BPS05,BCSW13}, and non-Newtonian and viscoelastic flows \cite{OCO06,MFD10,DNKS14} in geometrically-complex flow domains, with its potential applications in diverse fields extending from biomedical to the energy sector. The local nature of the calculations in the LB method makes the LB-based models amenable for code paralellization. Moreover, efficiencies in handling complex flow domain geometries and the explicit nature of the computational steps in the LB method facilitate its use in diverse applications. However, the computational efficiency of the LB-based numerical models need to be further improved to enhance their practical uses. 

Improvements on the computational performance of three modules of the LB-based numerical model coded in FORTRAN90 are reported in this paper; however, these optimization techniques are not limited to FORTRAN90 coding. The first module was developed to simulate the flow of multiple circular-cylinder micron-size particles in pressure-driven Newtonian fluid flow in a heterogeneous fractured porous domain formed by an array of non-overlapping, non-uniform stationary solid obstacles. The second module focused on the flow of nano-size particles that exhibit Brownian motion induced by thermal fluctuations in the fluid. The third module was developed to simulate non-Newtonian fluid (pseudoplastic and dilatant) flows in a confined channel. 

Several common LB performance enhancement techniques were initially applied to enhance the computational efficiency of the LB model. These techniques included the maximization of compile-time determinism, optimization of array memory layout for the streaming step, and OpenMP fine-tuning, which achieved 5-10$\times$ performance increases. Subsequently, additional techniques were applied to gain further performance increases, which involved generation of random numbers in a producer-consumer model and exploitation of task-level parallelism to perform task pipelining and output offloading. This yielded an additional 2-4$\times$ performance increase on simulations with thermal fluctuations and/or heavy I/O loads. Overall, an average of 21$\times$ computational performance increase was achieved through the combined application of these optimization techniques.


\section{Lattice-Boltzmann (LB) Model}
\label{sec:LBM}

In the lattice-Boltzmann (LB) method, the mesodynamics of the incompressible, Newtonian fluid flow is described by \cite{S01,W00,BSV92} with a single relaxation time via the continuous Bhatnagar-Gross-Krook (BKG) equation \cite{BGK54}

\begin{equation}
 \label{e.LB1} f_{i}\left(\mathbf{r+e}_{i}{\triangle t},t+{\triangle
 t}\right) -f_{i}\left(\mathbf{r},t \right) =\frac{\triangle t}{\tau} [
 {f_{i}^{eq}\left(\mathbf{r},t \right)-f_{i}\left(\mathbf{r},t
 \right) } ],
 \end{equation}

\noindent where $f_i(\mathbf{r},t)$ is the complete set of population densities of discrete velocities $\mathbf{e}_i$ at position $\bf{r}$ and discrete time $t$ with a time increment of $\triangle t$, $\tau$ is the relaxation parameter, and $f_i^{eq}$ is the local equilibrium \cite{QDL92}

 \begin{equation}
 \label{e.LB2}f_{i}^{eq}=\omega_i \rho \left
 (1+\frac{\mathbf{e}_i\mathbf{\cdot}\mathbf{u}}{c_s^2}
 +\frac{(\mathbf{e}_i\mathbf{\cdot}\mathbf{u})^2}{2c_s^4}-
 \frac{\mathbf{u \mathbf{\cdot}u}}{2c_s^2}\right),
\end{equation}

\noindent where $\omega_i$ is the weight associated with $\mathbf{e}_i$ 
and $c_s$ is the speed of sound, $c_s= \triangle x /\sqrt(3) \triangle t$. 
The local fluid density, $\rho$, and velocity, $\mathbf{u}$, at the 
lattice node are given by $\rho=\sum_{i} f_{i}$ and 
$\rho \mathbf{u}=\sum_{i} f_{i}\mathbf{e}_i+\tau \rho \mathbf{g}$, 
where $\mathbf{g}$ is the acceleration due to the force of gravity \cite{BG00}. 
The left hand side of Eq.~\ref{e.LB1} describes the streaming of population 
densities, $f_i$, from a lattice node $\mathbf{r}$ to the closest 
neighboring lattice node in the direction of $\mathbf{e}_i$ on a regular 
lattice grid. The right hand side of Eq.~\ref{e.LB1} describes the local 
collision process. A D2Q9 (two-dimensional nine velocity vector) 
lattice \cite{S01} was used in numerical simulations in this paper. 
For a D2Q9 model, 
$e_0=\left( 0,0 \right)$, $e_i=\left[ \cos \left( \frac{\left( i-1 \right) \pi }{2}\right),\sin \left( \frac{\left( i-1 \right) \pi }{2}\right) \right]$ 
for $i=1\mhyphen4$ (for the primitive lattice vectors of length unit), 
and $e_i=\sqrt{2}\left[ \cos \left( \frac{\left( 2i-9 \right) \pi }{4}\right),\sin \left( \frac{\left( 2i-9 \right) \pi }{4}\right) \right]$ for $i=5\mhyphen8$ 
(for diagonal lattice vectors of length $\sqrt{2}$), $\omega_0=\frac{4}{9}$, 
$\omega_i=\frac{1}{9}$ for $i=1\mhyphen4$ and $\omega_i=\frac{1}{36}$ for $i=5\mhyphen8$. 
Through the Chapman-Enskog expansion, the LB method for a single-phase 
flow recovers the Navier-Stokes equation for weakly compressible fluids

 \begin{equation}
 \label{NSE_1} \nabla \cdot \mathbf{u} =0,
\end{equation}
 \begin{equation}
 \label{NSE_2} \partial_{t} \mathbf{u} +\left( \mathbf{u} \cdot \nabla \right) \mathbf{u} = -\frac{\nabla p}{\rho}+ \nu \nabla^2 \mathbf{u} + \mathbf{g},
\end{equation}

\noindent with the fluid kinematic viscosity, $\nu= c_s^2 \triangle t \left( \tau - 0.5 \right) $. $\tau$ is computed as a function of $\nu$ and determines how fast population densities approach the equilibrium distributions upon collision in Eq.~\ref{e.LB1}. Pressure, $P$, is computed via the ideal gas relation, $P=c_s^2 \rho$. 

\section{Microparticle Lattice-Boltzmann Module}
\label{sec:model}

\begin{figure}[ht!]
\begin{center}
\includegraphics[width=0.8\textwidth]{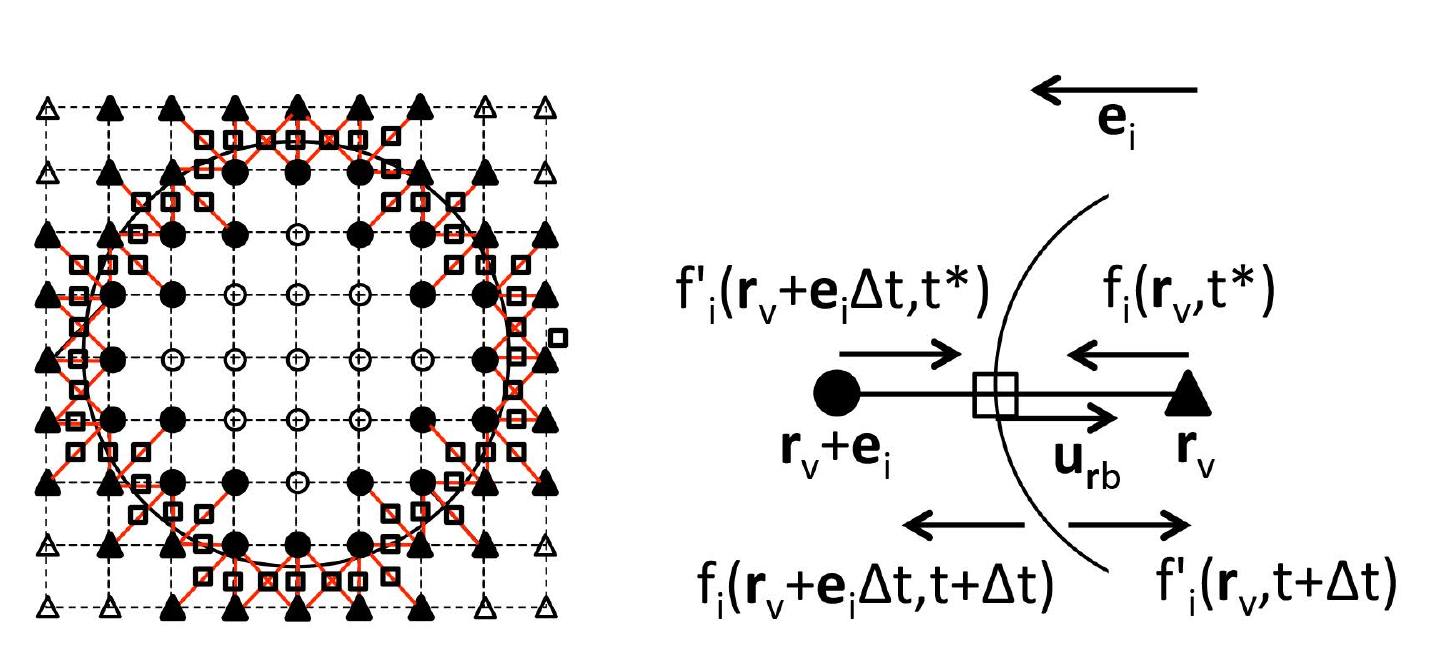}
\caption{Representation of a particle in the LB model and the momentum exchanges between a particle and the bulk fluid \cite{L94a,BMS08,BS10}. Filled circles inside the particle represent intra-particle virtual fluid nodes closest to the particle surface, filled triangles outside the particle represent extra-particle bulk fluid nodes, and the unfilled squares represent the boundary nodes located half-way between the intra- and extra-particle nodes.} \label{fig:LB_figures}
\end{center}
\end{figure}

The LB method described in Sec.~\ref{sec:LBM} was extended to simulate flow of micron-size particle in Newtonian fluids in low - moderate Reynolds number flows by accommodating particle-fluid hydrodynamics  \cite{BAS13,BMS08,BCF15}, based on model formulations in Ref. \cite{L94a,DA03,DL03,NL02}. In these formulations, the population densities near particle surfaces were modified to account for particle-fluid hydrodynamic forces,  $\mathbf{F}_{\mathbf{r}_b}$, which arise from momentum exchanges between the fluid outside the particle and the particle in motion \cite{L94a,DL03}. $\mathbf{F}_{\mathbf{r}_b}$ was calculated at particle boundary nodes located halfway between the intra-particle lattice node, $\mathbf{r}_v$, and extra-particle lattice node, $\mathbf{r}_v + \mathbf{e}_i$,  (Fig. \ref{fig:LB_figures})

\begin{figure}[ht!]
\includegraphics[width=0.55\textwidth]{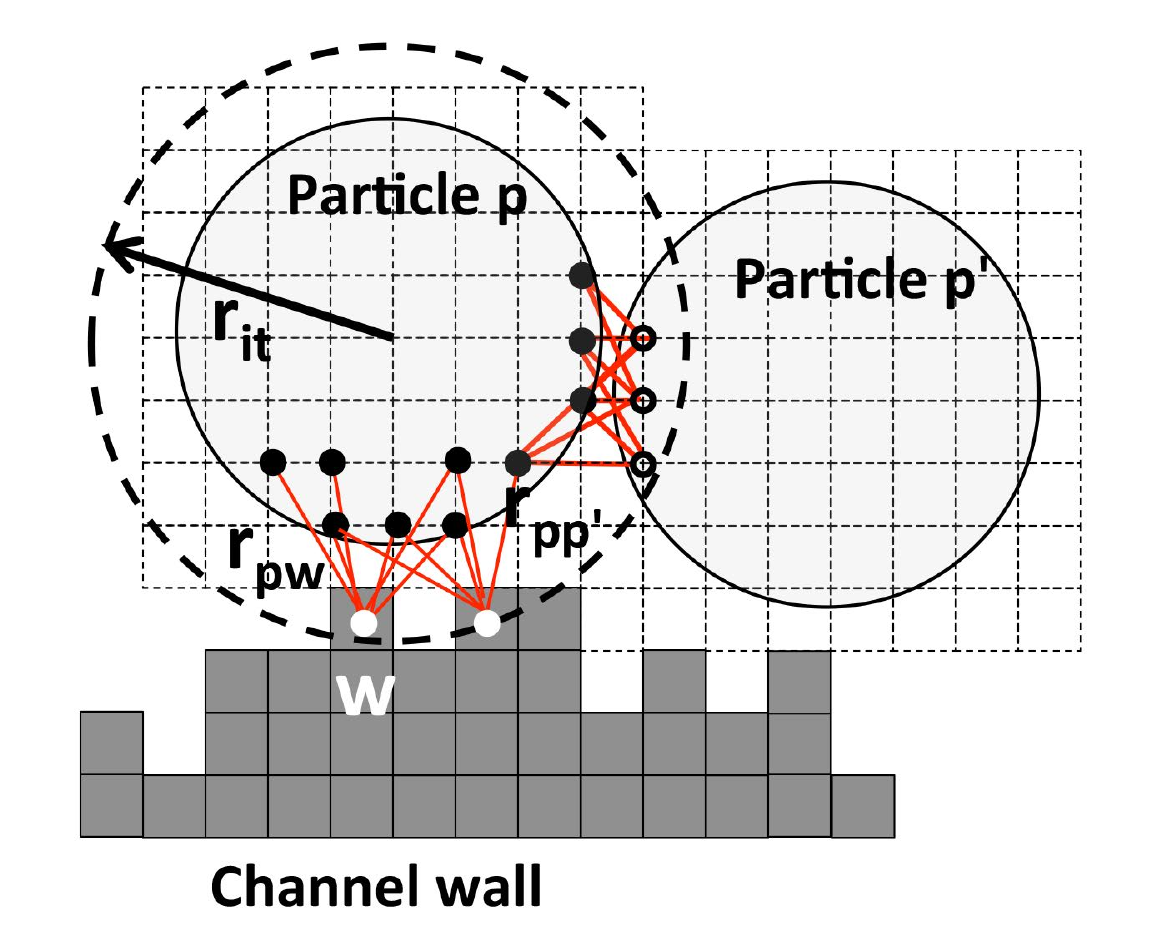}
\centering
\caption{Inter-particle and particle-wall repulsive interactions \cite{BS10}. Although a few interaction links (in red) are shown for simplicity, interaction potentials are defined over all intra-particle virtual nodes closest the particle surface (shown by black filled circles) and the wall nodes (denoted by w) and/or intra-particle virtual nodes of neighboring particles $p'$ (shown by open circles) within a threshold interaction distance, $r_{it}$, of a particle $p$.} \label{fig:LJ}
\end{figure}

 \begin{equation}
 \label{e.MP1} \mathbf{F}_{\mathbf{r}_b}=-2\left[f^{\prime}_i\left(\mathbf{r}_v+\mathbf{e}_i \triangle t,t^{\ast}\right)+
\frac{\rho\omega_i}{c_s^2}\left(\mathbf{u}_{\mathbf{r}_b} \cdot
\mathbf{e}_i\right)\right]\mathbf{e}_i,
 \end{equation}

\noindent
where $f^{\prime}_i$ is the population density in the $-\mathbf{e}_i$ direction at the post-collision time $t^{\ast}$, and $\mathbf{u}_{\mathbf{r}_b}$ is the local particle velocity at the boundary node $\mathbf{r}_b$.  Repulsive interaction forces between the particles and between the particles and stationary solid zones, including channel walls and inline obstacles, were expressed in terms of mathematically simplified form of two-body Lennard-Jones (LJ) potentials \cite{BS10}

 \begin{equation}
 \label{e.MP2} \mathbf{F}_{\mathbf{r}_{i}}=-\psi \left( \frac{\mid \mathbf{r}_i \mid} {\mid \mathbf{r}_{it} \mid}   \right)^{-13} \mathbf{n},
 \end{equation}

\noindent
where $|\mathbf{r}_i \mid$ is the distance between a particle surface node and the neighboring particle surface node ($\mathbf{r}_{i} = \mathbf{r}_{pp'}$) or between a particle surface node and the stationary solid node located on channel walls or inline obstacles ($\mathbf{r}_{i} = \mathbf{r}_{pw}$), $p$ is the particle index, $\mid \mathbf{r}_{it} \mid$ is the repulsive threshold distance, $\mathbf{n}$ is the unit vector along $\mathbf{r}_{i}$, and $\psi$ is the stiffness
parameter used to adjust the repulsive strength between the particles and between the particles and stationary solid zones (Fig. \ref{fig:LJ}). $\psi$ can be tuned to adjust repulsive interaction between neighboring particles or between a particle and stationary solid objects in near contact. The total hydrodynamic force, $\mathbf{F}_T$, exerted on the particle is

\begin{equation}
 \label{e.MP3}
 \mathbf{F}_T=\sum_{\mathbf{r}_b}
\mathbf{F}_{\mathbf{r}_b}+\sum_{\mathbf{r}^{c,u}_b}\mathbf{F}_{\mathbf{r}^{c,u}_b}
+\sum_{\mid \mathbf{r}_{pw} \mid \le \mid \mathbf{r}_{it} \mid} \mathbf{F}_{\mathbf{r}_{pw}}
+\sum_{\mid \mathbf{r}_{pp'} \mid \le \mid \mathbf{r}_{it} \mid} \mathbf{F}_{\mathbf{r}_{pp'}},
\end{equation}

\noindent
where $\mathbf{F}_{\mathbf{r}^{c,u}_b}=\pm \rho
\left(\mathbf{u}_{\mathbf{r}_b^{c,u}}- \mathbf{U}_p \right)/ \triangle t$
is the additional force induced by covered, $\mathbf{r}^c_b$, and
uncovered, $\mathbf{r}^u_b$, lattice nodes due to particle motion
\cite{DA03,BMS08,DL03}, and $\mathbf{U}_p$ is
the particle translational velocity.  In Eq.~\ref{e.MP3}, it was assumed that the fluid occupies the entire flow domain to ensure continuity in the flow field to avoid large artificial pressure gradients that may arise from compression and expansion of the fluid near the particle surface. Momentum exchanges occur between the mobile particle and the bulk fluid, following the method of Nguyen and Ladd \cite{NL02}. The total torque on the particle, $\mathbf{T}_T$ is defined as

\begin{eqnarray}
 \label{e.MP4}
\mathbf{T}_T &=& \sum_{\mathbf{r}_b} \left( \mathbf{r}_b - \mathbf{r}_c
\right)\times \mathbf{F}_{\mathbf{r}_b}+\sum_{\mathbf{r}^{c,u}_b}
\left( \mathbf{r}^{c,u}_b - \mathbf{r}_c \right) \times
\mathbf{F}_{\mathbf{r}^{c,u}_b}\nonumber
+ \sum_{\mid \mathbf{r}_{pw} \mid \le \mid \mathbf{r}_{it} \mid} \left( \mathbf{r}_{pw} - \mathbf{r}_c \right) \times \mathbf{F}_{\mathbf{r}_{pw}} \\
&+& \sum_{\mid \mathbf{r}_{pp'} \mid \le \mid \mathbf{r}_{it} \mid} \left( \mathbf{r}_{pp'} - \mathbf{r}_c \right) \times \mathbf{F}_{\mathbf{r}_{pp'}}.
 \end{eqnarray}

The translational velocity, $\mathbf{U}_p$, and the angular velocity of the particle, $\mathbf{\Omega}_p$, are advanced in time according to the discretized Newton's equations of motion 

\begin{equation}
 \label{e.MP5}
\mathbf{U}_p\left( t+\triangle t\right) \equiv
\mathbf{U}_p\left(t\right)+\frac{\triangle t}{m_p}\mathbf{F}_T
\left( t\right)+\frac{\triangle t}{\rho_p}(\rho_p-\rho)\mathbf{g},
 \end{equation}

\begin{equation}
 \label{e.MP6} \mathbf{\Omega}_p\left( t+\triangle t\right) \equiv
 \mathbf{\Omega}_p\left(t\right)+\frac{\triangle t}{I_p}\mathbf{T}_T\left( t\right), 
 \end{equation}

 \noindent  where $m_p$ is the particle mass and $\mathbf{u}_b=\mathbf{U}_p+\mathbf{\Omega}_p\times\left({\mathbf{r}_b} 
- \mathbf{r}_c \right)$. The new position of the particle is
computed as $\mathbf{r}_c\left(t+\triangle t \right)
=\mathbf{r}_c \left( t\right)+\mathbf{U}_p\left(t \right)\triangle
t$. The population densities at $\mathbf{r}_v$ and $\mathbf{r}_v
+\mathbf{e}_i \triangle t$ are updated to account for momentum-exchange between the particle and bulk fluid in accordance with \cite{L94a}

\begin{equation}
 \label{e.MP7} f^{\prime}_i\left(\mathbf{r}_v,t+\triangle t\right)=f_i(\mathbf{r}_v,t^{\ast})-\frac{2\rho \omega_i}{c_s^2}\left(\mathbf{u}_{\mathbf{r}_b} \cdot \mathbf{e}_i \right),
 \end{equation}

 \begin{equation}
 \label{e.MP8} f_i\left(\mathbf{r}_v+\mathbf{e}_i \triangle t,t+\triangle t\right)=f^{\prime}_i(\mathbf{r}_v+\mathbf{e}_i \triangle t,t^{\ast})+\frac{2\rho
\omega_i}{c_s^2}\left(\mathbf{u}_{\mathbf{r}_b}  \cdot \mathbf{e}_i \right).
 \end{equation}

The microparticle LB model successfully simulated particle trajectories and velocities of micron-size particles in a microfluidic channel with staggered inline obstacles \cite{BAS13}. The model also successfully captured transition in flow migration modes ranging from steady equilibrium with a monotonic approach to strong oscillatory motion in a 2D smooth-walled flow channel with gradual increases in the flow Reynolds number \cite{BAS13,BMS08}, consistent with the previously published numerical simulations based on the finite-element method \cite{FHJ94}.    

\section{Fluctuating LB-BGK (FLB-BKG) Module}
\label{sec:FLB}

If particles are of submicron-size, particle Brownian motion could affect particle trajectories. Particle kinetic energy associated with its Brownian motion should be in balance with the thermal energy of the fluid for numerical simulations to obey the fluctuation-dissipation theorem. Random noise (thermal fluctuations) introduced into the fluid randomly perturbs particle velocity as the particle exchanges momentum with the fluid.  In the FLB-BGK formulation, thermal fluctuations are added to the post-collisional populations of the fluid \cite{ASC05}

\begin{equation}
 \label{e.NP1}
 f^{\ast}_i\left(\mathbf{r},t^{\ast}\right)=f_i\left(\mathbf{r},t^{\ast}\right)+\triangle
 t \zeta_i \left(\mathbf{r},t\right),
 \end{equation}

\noindent
where $f_i\left(\mathbf{r},t^{\ast}\right)$ is the post-collisional
population density and $\zeta_i \left(\mathbf{r},t\right)$ is
space/time local distribution and acts at the level of the stress
tensor and non-hydrodynamic modes, which can be written as \cite{BMSY12}

\begin{equation}
 \label{e.NP2}
 \zeta_i \left(\mathbf{r},t\right)=\sqrt{\frac{\rho k_B T \varpi \left( 2-\varpi \right)}{c_s^2}}
 \sum_{k=3}^{8} \omega_i \chi_{ki} \Theta_k,
 \end{equation}

\noindent
where $k_B$ is the Boltzmann constant, $T$ is the temperature, $\varpi=\triangle t / \tau$ for a Newtonian fluid. $\left\{\chi_{ki} \right\}_{k=0,8}$ is a set of nine lattice eigenvectors mutually orthonormal according to the scalar products $\sum_{i=0}^{8} \omega_i \chi_{ki} \chi_{li}=\delta_{kl}$ and $\sum_{k=0}^{8} \omega_i \chi_{ki} \chi_{ki^\prime}=\delta_{ii^\prime}$, and $\Theta_k$ is a set of six normal deviates. The eigenvectors for the D2Q9 lattice model correspond to the kinetic moments,
 in which $k$=0 is the mass density, $k$=1-2 components of the momentum, $k$=3-5 components of the
 deviatoric momentum flux, and the remaining $k$=6-8 eigenvectors correspond to the
  ghost modes \cite{ASC05,AS08}.  These eigenvectors can be expressed as $\chi_{ki}= {e_{ki}}/{\sqrt{w_k}} $,
  in which the vectors $e_{ki}$ and their length, ${w_k}$, are listed in Table ~\ref{t1}.

\begin{table}[h]
  \caption{Basis vectors of the D2Q9 model \cite{DSL07}. ${w_k}$ is the length of the $k$th basis vector given by
  ${w_k}=\sum_i {\omega_i e_{ki}^2}$ .}
\begin{center}
\scalebox{0.80} {
 \begin{tabular}{c|c|c|c|c|c|c|c|c|c r} 
 \hline  
 $k$ & 0 & 1 & 2 & 3 & 4 & 5 & 6 & 7 & 8    \\
 \hline \hline
 $e_{ki}$ & 1 & $\hat{e}_{ix}$ & $\hat{e}_{iy}$ & $3\hat{e}_{i}^2-2$ & $2\hat{e}_{ix}^2-\hat{e}_{i}^2$ & $\hat{e}_{ix}\hat{e}_{iy}$ & $\left(3\hat{e}_i^2 -4 \right)\hat{e}_{ix} $ & $\left(3\hat{e}_i^2 -4 \right)\hat{e}_{iy} $ & $9\hat{e}_i^4-15\hat{e}_i^2+2$                   \\

$w_k$ & 1 & 1/3 & 1/3 & 4 & 4/9 & 1/9 & 2/3 & 2/3 & 16    \\

 \hline
 \end{tabular}
 }
 \end{center}
 \label{t1}
  \end{table}

Using the FLB model, we previously demonstrated (i) perfect thermalization of a fluctuating fluid in a confined channel in the presence of a suspended particle near the channel wall under no external force; (ii) a crossover from a ballistic regime to a diffusive regime at which particle velocity autocorrelation vanished while the ratio of the mean-squared displacement of particle positions to the elapsed time approached unity in the diffusive regime; and (iii) the particle obeyed the fluctuation-dissipation theorem (FDT) in a flowing fluid, when it is not in the vicinity of the walls and/or inertial forces are not large \cite{BMSY12}.

\section{Non-Newtonian Fluid Flow Module} \label{NNFS}

The fluid flow in the LB methods described in Sections~\ref{sec:LBM} $\mhyphen$~\ref{sec:FLB} is Newtonian. Non-Newtonian fluid flow can be simulated by accommodating the nonlinear relations between the rate of strain and stress rate of the fluid, through which the local fluid kinematic viscosity dynamically evolves in time and space as a function of local velocity gradients. 

Numerical modeling of early experimental studies of non-Newtonian fluids in pipe flows measuring the pressure gradient versus bulk fluid velocity for a wide range of fluids \cite{DM59} suggested a power law relation between fluid viscosity and shear rate, $\dot \gamma$. Such power law relations can be built on, for example, via the Ostwald-de Waele model (Eq.~\ref{nn1}) 

 \begin{equation}
 \label{nn1} \eta= \xi \dot \gamma  ^ {n-1}       .
 \end{equation}

\noindent
where $\eta$ is the dynamic viscosity of the fluid ($\eta=\rho \nu$), $\xi$ is the consistency, $n$ is the fluid-type identifier, $n<1$, $n=1$, and $n>1$ correspond to pseudoplastic (shear-thinning), Newtonian, and dilatant (shear-thickening) fluids, respectively. The shear rate, ${\dot \gamma}$ can be computed from the the second invariant of the rate of strain tensor, $\Pi_{D}$ \cite{NAV11}  

\begin{equation}
 \label{nn2} \dot \gamma= 2 \sqrt{\Pi_{D}}.
 \end{equation}

 \begin{equation}
 \label{nn3} 
 \begin{aligned}
 \Pi_D = \frac{1}{4} \left[ \left( \frac{\partial u}{\partial y}+ \frac{\partial v}{\partial x} \right) \right]^2 - \frac{\partial u}{\partial x} \frac{\partial v}{\partial y}  ,
  \end{aligned}
 \end{equation}

\noindent
Eq.~\ref{nn3} was obtained from $\Pi_{D}=\frac{1}{2} \left(  \left[ tr\left(\mathbf{D} \right) \right]^2  - tr(\mathbf{D}^2) \right)$, where $\mathbf{D}=\frac{1}{2} \left( \nabla \mathbf{u}(\mathbf{x})+  (\nabla \mathbf{u}(\mathbf{x)})^T \right)$, $\mathbf{u}=\left(u,v\right)$, $\mathbf{x}=\left(x,y \right)$, and $T$ is the transpose operator. From  Eqs.~\ref{nn1} and ~\ref{nn2} for incompressible fluids ($\rho=1$ in lattice units),  

\begin{equation}
 \label{nn4} \nu^{*}= \left[ \kappa   {| \Pi_{D} |}    ^  {\frac{n-1}{2}} \right]  \xi .
 \end{equation}

\noindent
Different values were used for $\kappa$ in Eq.~\ref{nn4} in the earlier numerical models based on the LB method. For example, $\kappa = 0.5^{n-1}$ in \cite{PKYPK07}, $\kappa = 1$ in \cite{HR11}, and $\kappa = 2^{n-1}$ \cite{DNKS14}. Our numerical simulation indicated that these $\kappa$ values performed equally-well in calculating normalized steady fluid velocities in a smooth-walled flow channel. We adopted $\kappa = 2^{n-1}$ in the simulations used to test the computational performance enhancements in this paper. 

\begin{figure}[ht!]
\begin{center}
\includegraphics[width=0.65\textwidth]{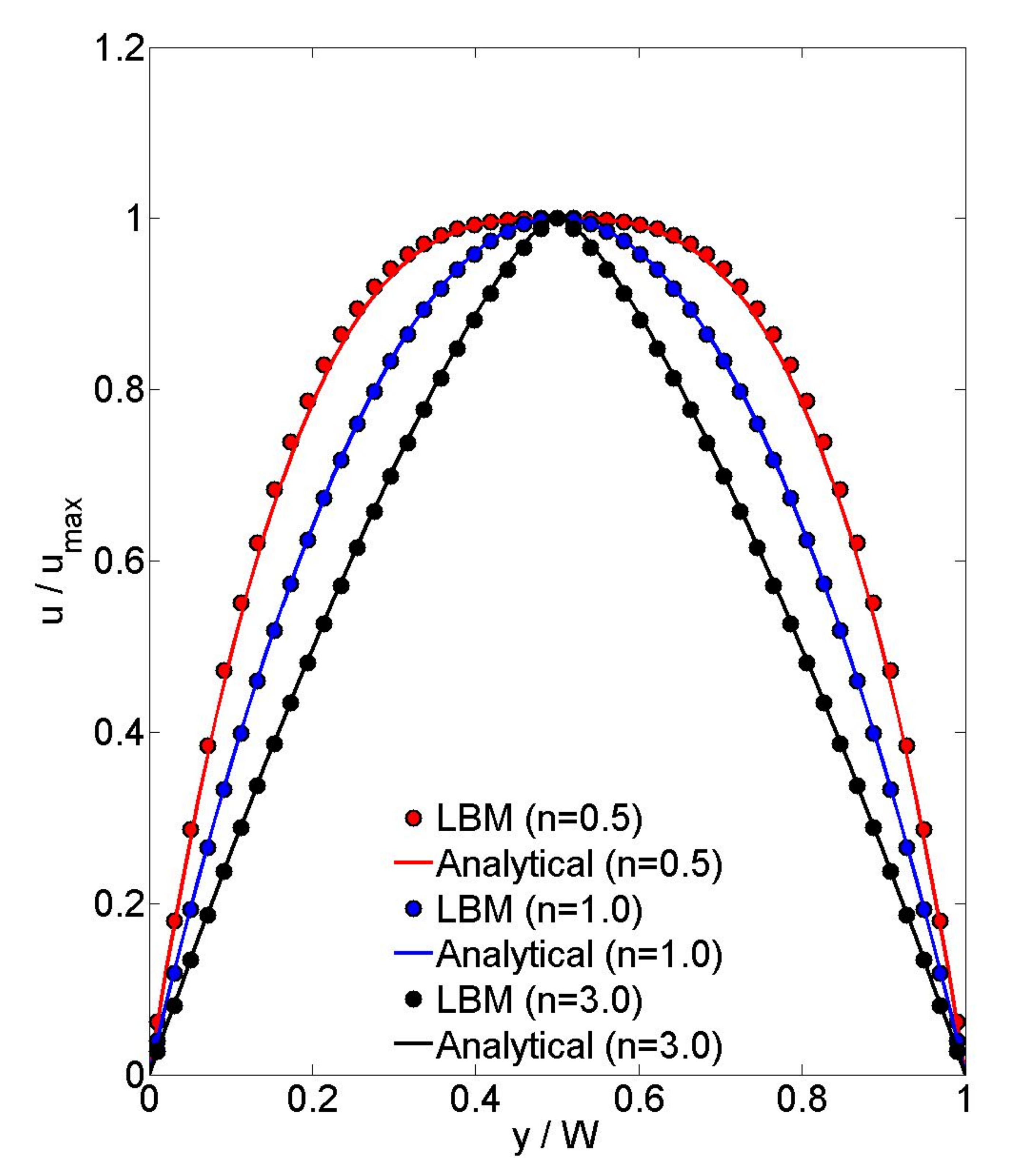}
\caption{Normalized steady velocity profiles of gravity-driven pseudoplastic $(n=0.5)$, Newtonian $(n=1)$, and dilatant $(n=3)$ fluids in a horizontal, smooth-walled flow channel for $\kappa = 2^{n-1}$.  $\xi$ was set to $10^{-3}$.  Fluid velocities were normalized with respect to the maximum fluid velocity, $u_{max}$. LBM stands for the lattice-Boltzmann model simulations.} \label{fig:valid}
\end{center}
\end{figure}

In Eq.~\ref{nn2}, $\nu^{*} \to \infty$ for $n<1$ and $\nu^{*} \to 0$ for $n>1$ as $\Pi_{D} \to 0$, which are both unphysical. To avoid unphysical values for $\nu^{*}$, the lower and upper bounds for $\nu^{*}$ can be obtained by setting the relaxation parameter in the LBM $\tau \sim 0.5$ (to ensure numerical stability) and $\tau \sim 1.0$ (to ensure enhanced accuracy) \cite{GDK05}. In this formulation, the local kinematic viscosity, $\nu^{*}$ would make the relaxation parameter, $\tau^*$ local, through $\tau^{*}=0.5+3\nu^{*}  ( \triangle t / \triangle x^2  )$, where $\nu^{*}$ is introduced by Eq.~\ref{nn4}. As a result, Eq.~\ref{e.LB1} for non-Newtonian fluid flow becomes

 \begin{equation}
  \label{nn5} f_{i}\left(\mathbf{r+e}_{i}{\triangle t},t+{\triangle
  t}\right) -f_{i}\left(\mathbf{r},t \right) =\frac{\triangle t}{\tau^{*}} [
  {f_{i}^{eq}\left(\mathbf{r},t \right)-f_{i}\left(\mathbf{r},t
  \right) } ].
  \end{equation}

\noindent
and the momentum equation at each lattice node is computed through $\rho \mathbf{u}=f_{i}\mathbf{e}_i+\tau^{*} \rho \mathbf{g}$.  The generalized analytic solution for the steady-state velocity profile of non-Newtonian fluid flows in a smooth-walled horizontal channel, is given by \cite{PKYPK07,W68}:

 \begin{equation}
 \label{nn6} u(y) = u_o \left[ 1- \left(  \frac{2|y|}{W}  \right)^{1+\frac{1}{n}}  \right],
 \end{equation}

 \begin{equation}
 \label{nn7} u_o =  \left( \frac{1}{\xi} |\mathbf{g}| \right)^{\frac{1}{n}}  \left(  \frac{W}{2}\right)^{1+\frac{1}{n}} \left( \frac{n}{n+1} \right) ,
 \end{equation}

\noindent
where $W$ is the channel width and $y$ is vertical distance (perpendicular to the main flow direction) from one of the channel walls. The upper and lower limits for $\nu^*$ were set to $10^{-5}$ and $0.1$ to prevent unphysical values for $\nu^*$.  A successful validation of the non-Newtonian module against the analytic solution given in Eqs.~\ref{nn6}~-~\ref{nn7} is shown in Fig. \ref{fig:valid}.

\pagebreak

\section{Benchmark Problems and Numerical Simulations}
\label{sec:sims}

Three benchmark problems were set up to measure the computational efficiency of the combined implementation, when applicable, with the optimization techniques proposed in this paper. These benchmark problems involve (i) flow of initially closely-packed multiple particles in a Newtonian fluid flow driven by diagonally-aligned external force in a heterogeneous and disordered fractured-granular porous medium; (ii) nanoparticle flow and thermal fluctuation in the background fluid in a smooth-walled channel, and (iii) non-Newtonian fluid flow in a horizontal channel. 

We report all the LB input parameters and computed variables in lattice units in the subsequent section to facilitate ease of repeatability of the numerical simulations and results. $\triangle x= \triangle t =1$ was imposed in all simulations discussed in this section. Translation of LB parameters from lattice units to physical dimensions, however, can be found in Refs. \cite{S01, BAS13}. 

 \vspace*{0.2cm}
\noindent
\textbf{\textit{Benchmark 1: Multi-particle Flows in Heterogeneous Porous Media}}
 \vspace*{0.2cm}

\begin{figure}[ht!]
\begin{center}
\includegraphics[width=1.2\textwidth]{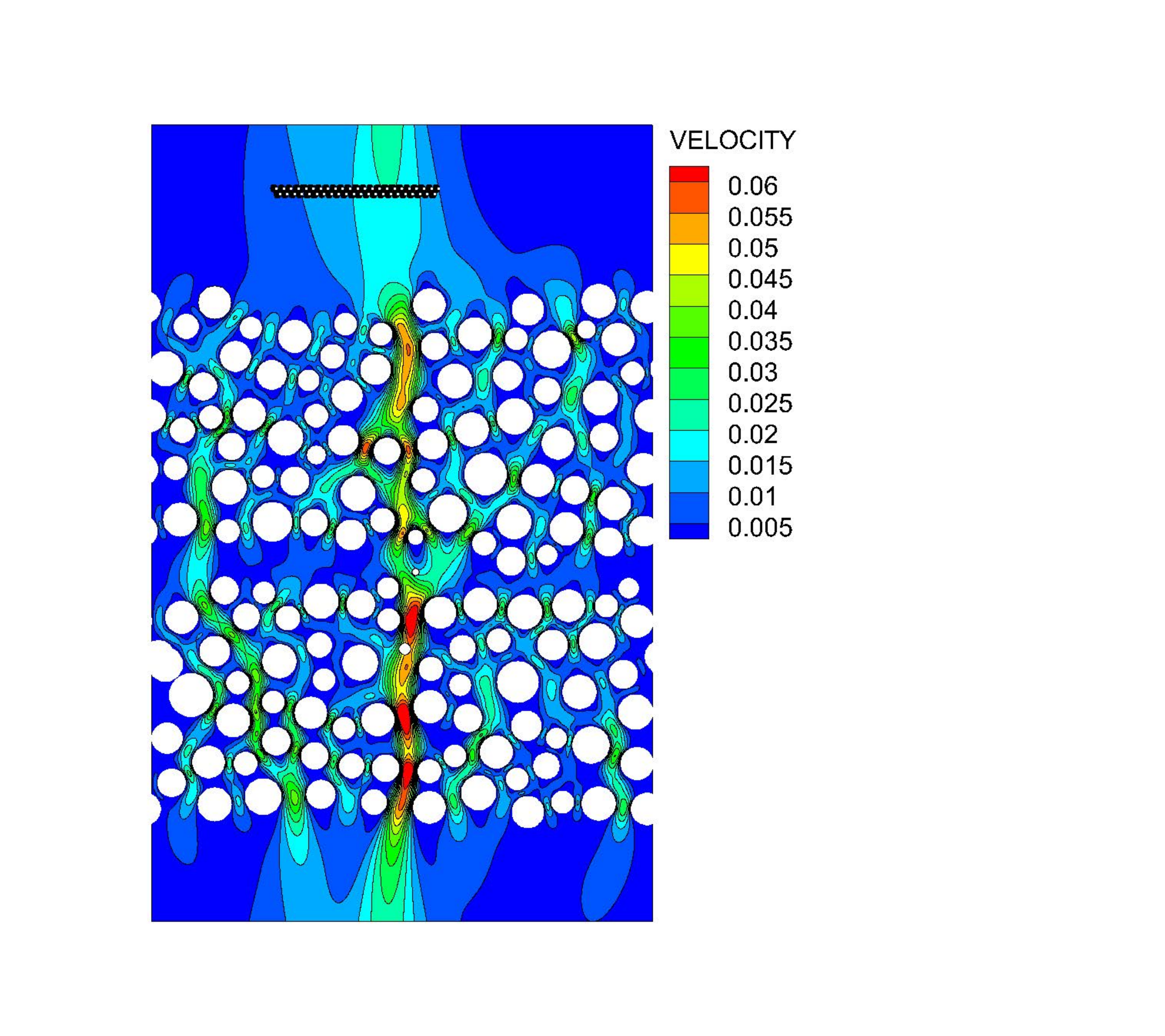}
\caption{Release location of a total of 45 cylindrical-circular particle into a steady flow field in a heterogeneous, disordered, fractured granular-porous domain. Contours represent the flow velocity field, whose magnitudes are expressed in lattice units. Particles are shown by black-filled circles. Stationary solid grains are shown by white-filled circles.} \label{fig:steady_porous}
\end{center}
\end{figure}

A numerical simulation was set up to simulate flow trajectories and velocities of cylindrical-circular particles in a spatially-heterogeneous, disordered fractured-porous media (Fig. \ref{fig:steady_porous}). The fluid flow was assumed to be Newtonian, which was driven by a diagonally-oriented external force with a force strength of $|\mathbf{g}|=(10^{-6}, 10^{-5})$. No-flow condition was imposed along the lateral boundaries and periodic boundaries were implemented at the inlet and outlet. The domain size was set to $1,020\times1,620$. The kinematic viscosity of the fluid was set to to 0.136, which results in $\tau=0.636 < 1$ that met the numerical stability requirement of the LB method.

\begin{figure}[ht!]
\begin{center}
\includegraphics[width=1.2\textwidth] {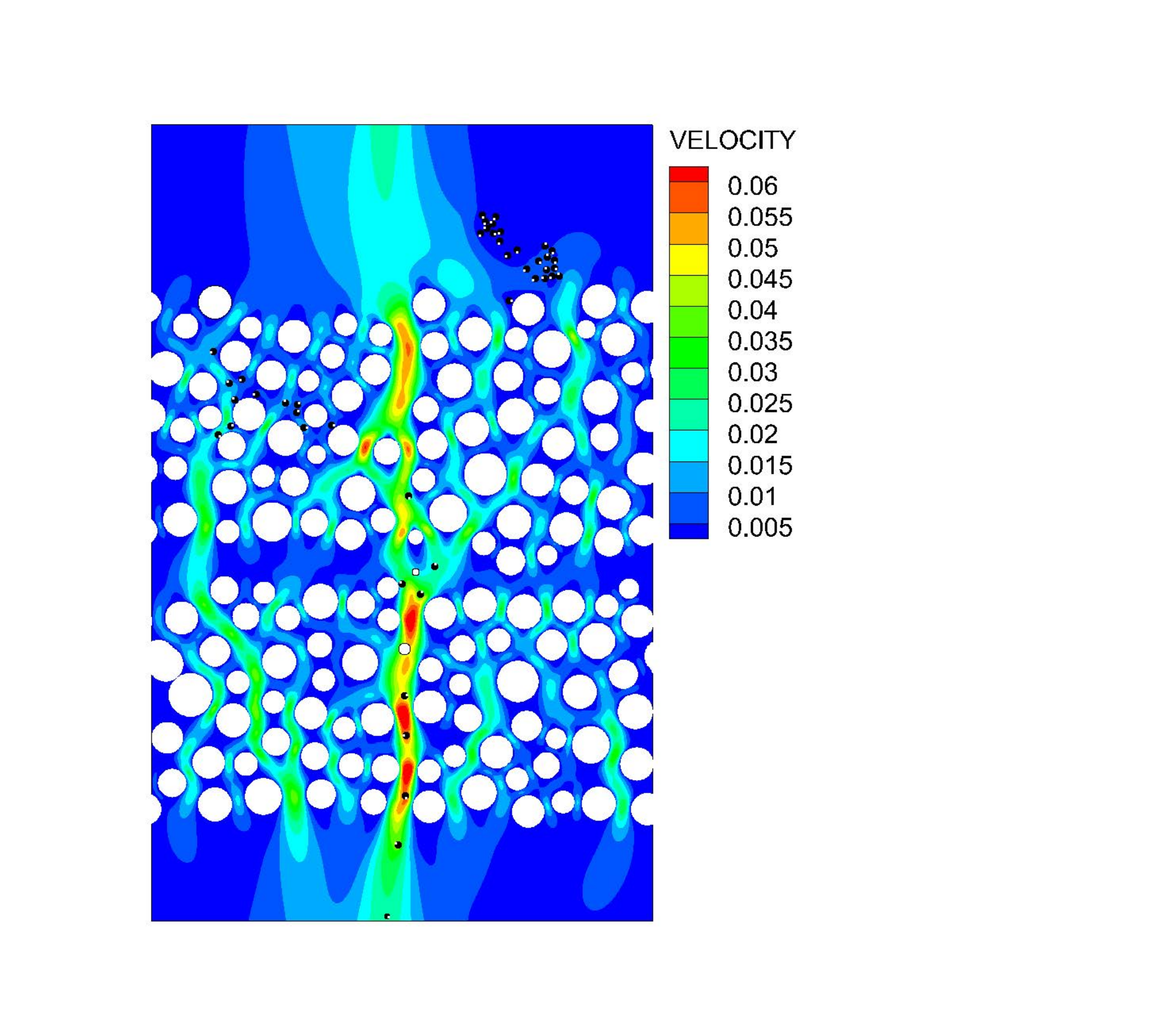}
\caption{Final position of cylindrical-circular particles. Contours represent the flow velocity field, expressed in lattice units. Particles are shown by black-filled circles. White dots on the particles are used to show the angular rotation of the particles. } \label{fig:final_porous}
\end{center}
\end{figure}

Once the steady-state flow field was obtained within 1$\%$, 45 closely-packed cylindrical-circular particles were released into the flow field near the inlet. The particle radius was set to 6.5 lattice-spacing. Their release locations are shown in Fig. \ref{fig:steady_porous}, in which surface-to-surface horizontal spacing between neighboring particles was set to 8.5 and the vertical surface-to-surface spacing between the top and bottom arrays of particles was set to 5.5. The LJ repulsive force, Eq.~\ref{e.MP2}, was implemented between particles as well as between a particle and the closest wall with repulsive strength of 20, if the separation distance between them is less than or equal to 2.5. Final positions of the particles are shown in Fig. \ref{fig:final_porous}. The simulation ended when one of the particles reached the lower boundary, corresponding to $5.6\times10^4$ time-step after steady-flow field was established.

 \vspace*{0.2cm}
\noindent
\textbf{\textit{Benchmark 2: Submicron-size Particle Flow Simulations}}
 \vspace*{0.2cm}

\begin{figure}[ht!]
\begin{center}
\includegraphics[width=0.8\textwidth] {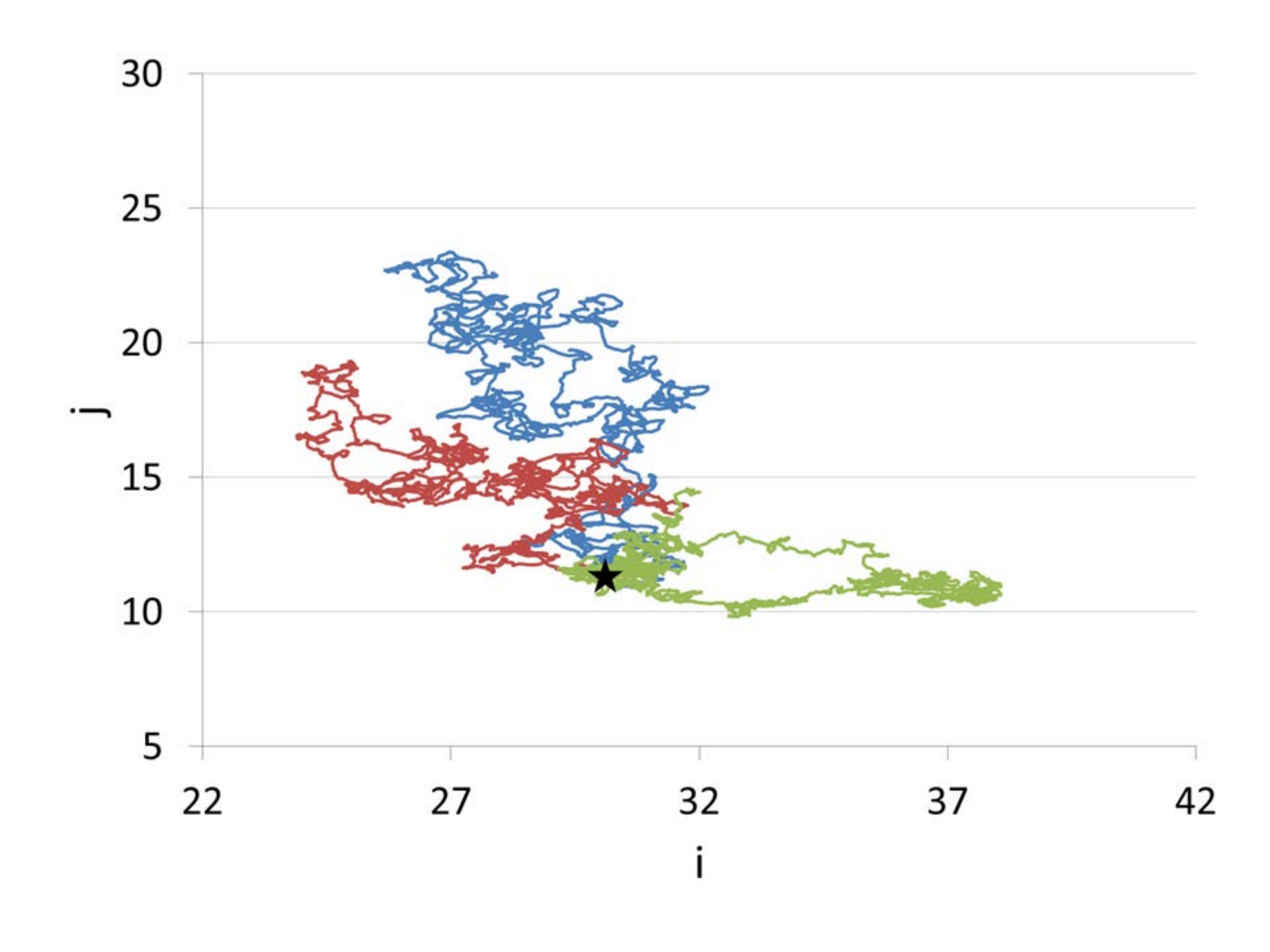}
\caption{Three realizations of trajectories of a Brownian particle released into an initially stagnant Newtonian fluid in a narrow smooth-walled channel. Star marks the release location of the particle and (i,j) are the lattice indices.} \label{fig:flb}
\end{center}
\end{figure}

For submicron-size particle simulations via the FLB, a lattice domain of 3,000
$\times$41 was used. External force strength, $\mathbf{g}$, 
was set to zero so that thermal fluctuations in the fluid were the only 
driving force for the particle motion. A single particle of radius 7.5 was 
placed into a horizontal channel and simulated for $8\times10^5$ time-step. 
$k_{B}T$ and $\nu$ were set to $0.00005$ and 0.1, respectively. 
Of the 28 available cores, 24 were used as OpenMP threads, and 4 dedicated 
to random number generation. Simulated trajectories of a Brownian particle 
released into Newtonian fluid in a confined channel are shown in Fig. \ref{fig:flb}. 
In these simulations, the fluid was assumed to be initially stagnant. 
Particle motion is due to combined effects of thermal fluctuations-induced 
particle Brownian motion (Eq.~\ref{e.NP1}), particle-fluid hydrodynamics (Eq.~\ref{e.MP1}), 
and the wall effects. 

\pagebreak

 \vspace*{0.2cm}
\noindent
\textbf{\textit{Benchmark 3: Non-Newtonian Flow in a Smooth-Walled Channel}}
 \vspace*{0.2cm}

A benchmark problem that calculated the steady-state velocity profile of a pseudoplastic fluid flow with $n=0.5$ in a smooth-walled horizontal channel (Fig. \ref{fig:valid}) was used. The domain size was set to $101 \times 51$, and  $\xi$ was set to $10^{-3}$.  $2\times 10^{6}$ time-step was used to obtain steady-state velocity profile of the non-Newtonian fluid flow.

\section{Computational Enhancement Methods}
\label{sec:seql}

Most computational enhancements to LB models typically focus on algorithmic optimizations, such as what data structures were used to hold the fluid/solid nodes. In this work, we extended this view with a strictly  computational perspective to achieve significant performance improvements. We benchmarked our simulations on a 14 physical core/28 virtual core (via hyper-threading) Intel(R) Xeon(R) CPU E5-2697 v3 2.6 GHz quad-processor machine with 64 GB RAM running Ubuntu 14.04. We used the Intel Fortran compiler v16. The model performance is shown in Table.~\ref{table:t8}, in which MLUPS (mega lattice-site updates per second) was computed via MLUPS$=(K_x K_y)/(10^6 T)$, where $K_i$ is the number of grid points in the $i$th direction and $T$ is the time taken to update the entire lattice grid \cite{SSKDKSA13}.

\begin {table}[h!]
\caption{MLUPS for the original and optimized code. Increasing computational performance is shown from left to right as additional optimization techniques were applied. RNG represents MLUPS due to moving random number generation from a single thread to a producer-consumer model. KMP\_AFFINITY corresponds to the MLUPS due to nuanced OpenMP thread placement to maximize data locality.}
\centering
 \begin{tabular}{c|c|c|c|c|c r} 
 \hline
 Benchmark & Original  & RNG & Serial Optimizations & OpenMP & \verb|KMP_AFFINITY|   \\ 
 \hline\hline
 1 & 2.82 & - & 6.62 & 37.27 & 47.27 \\ 
 \hline
 2 & 3.42  & 5.76  & 15.00 & 82.75 & 138.20 \\
 \hline
3 & 5.89 & - & 12.26  & 57.10 & 85.85 \\ 
 \hline
\end{tabular}
\label{table:t8}
\end {table}

The following optimization techniques were used to enhance the computational performance of the model:

 \vspace*{0.2cm}
\noindent
\textbf{\textit{Compiler Flags}}
 \vspace*{0.2cm}

Through examining the compiler, we found that the default suggested flags for `fast' code (\verb|-O3 -fast|)
greatly reduced performance when combined with \verb|-openmp|, because \verb|-fast| sets \verb|-xHost|. For modern machines, this means utilizing the AVX2 vector register set. If these vector register sets are used in any OpenMP loops, then the maximum number of active threads is limited by the number of available sets of AVX registers. We also found that \texttt{-O3 -ipo -openmp -fp-model fast=2 -mcmodel=medium} produced the greatest performance increases across all simulations. Additional computational improvements were achieved by examining the optimization report provided by the compiler, addressing areas where it was not able to vectorize because of unknown loop bounds (i.e., variable or run-time constant), sub-optimal loop orderings, and assumed vector/parallel dependencies. Consequently, we made all simulation-constant LB elements, such as the unit velocity vector ($\mathbf{e}_i$), weights ($\omega_i$), geometric dimensions, and mathematical expressions compile-time constants. This required the use of FORTRAN03 language constructs, because initialization expressions and literal constants are not supported in FORTRAN95. This substantially increased the compiler's ability to optimize, vectorize, and unroll complex execution flow. Overall we achieved a 2-4$\times$ serial performance increase through manipulation of compiler invocation and maximization of compile-time determinism, when compared with the un-optimized code (Table~\ref{table:t8}). While this technique was only applied to 2D simulations, we believe it will provide even greater performance increases when applied to 3D simulations, because the memory requirements for 3D simulations are much greater.

 \noindent
\textbf{\textit{Improved Performance through Data Reduction}}
 \vspace*{0.2cm}

Through memory bandwidth analysis, we determined that the large arrays used in our LB code for maintaining ($u_x$,$u_y$) in non-Newtonian fluid flow simulations dramatically decreased performance. This was due to the increased cache pressure on the small, fast L1 and L2 CPU caches, resulting in large numbers of cache misses for data read/writes, and consequent increased usage of the slower main memory (RAM) to service those read/write requests \cite{MA02}. As the $u_x/u_y$ properties of a fluid node are independent of the $u_x/u_y$ values of its neighbors, we were able to eliminate the large (X $\times$ Y) array for maintaining this property, and use scalars instead. We achieved an additional performance increase of up to 10$\times$ from application of this technique due to decreased cache pressure and increased loop fusion/vectorization. However, this array minimization approach cannot be applied unilaterally, as we observed cases in which switching from large arrays to a few scalars minimally increases or even decreases performance. This can occur in loops featuring heavy floating point and vector register usage (such as during the collision step in the LB model), where reserving otherwise available registers to hold scalar subroutine results hinders the compiler optimization capabilities.

 \vspace*{0.2cm}
\noindent
\textbf{\textit{Random Number Generation for FLB-BGK model}}
 \vspace*{0.2cm}

The addition of thermal fluctuations to the LB-BGK model for submicron-size particle flow simulations required the generation of large chunks of random numbers each timestep. Our initial approach was to generate the random numbers in a large chunk immediately prior to usage (single thread); this had the side effect of constantly flushing large amounts of hot data from the caches. We addressed this computational bottleneck by shifting our generation strategy to a producer-consumer model, in which a C++ wrapper generates random numbers in large chunks in dedicated threads (separate from simulation thread(s)), and places them into a queue for consumption by the simulation. Each timestep, the simulation would consume one chunk of random numbers from the queue, greatly improving the cache hit/miss ratio. The effect of this technique is shown in Table ~\ref{table:t8} as part of the optimized parallel execution time for Benchmark 2, where the greatest overall increase in MLUPS was achieved. Table ~\ref{table:t8} reveals that moving the random number generator to separate threads prior to implementation of other optimization techniques resulted in 1.68$\times $ speed-up.

 \vspace*{0.2cm}
\noindent
\textbf{\textit{Code Parallelization via OpenMP}}
 \vspace*{0.2cm}

We utilized the Intel OpenMP runtime environment for distributing the computational workload across the cores of a system, per standard LB optimization. However, our use of compile-time determinism discussed above enabled the use of the \verb|$omp schedule(static)| directive for loop parallelization, greatly reducing run-time overhead in comparison to \verb|$omp schedule(dynamic)| that is mostly commonly applied. We further optimized the performance of OpenMP threads by disallowing floating between cores by setting

\verb|KMP_AFFINITY=compact,granularity=fine,1,0|. 

In a dual processor system, for example, threads 0 and 1 (which would likely be operating on adjacent sections of an array) are precluded from being assigned to separate processors, which greatly reduces inter-processor communication. Such communication is several orders of magnitude slower than intra-processor core communication. Overall, average performance increased by 8$\times$ beyond the optimized serial code, as shown in Table ~\ref{table:t8}. In our simulations, the collision/streaming steps constituted the bulk of the computational workload, even after the particles were released; therefore, we decided not to parallelize the particle-particle or particle-wall interactions beyond the auto-parallelization performed by the compiler.

\begin{table}[h!]
\caption {Simulation speed-ups.}
\centering
 \begin{tabular}{c|c|c|c r} 
 \hline
Benchmark & Serial Speed-up & Parallel Speed-up & Overall Speed-up \\
 \hline\hline
1 &  2.34  & 7.15  & 21.77 \\ 
 \hline
2& 4.38 & 9.21 & 40.41\\
 \hline
3  & 2.08  & 7.0 & 14.58 \\ [1ex]
 \hline
\end{tabular}

\label {table:t9}
\end{table}

\vspace*{0.2cm}
\noindent
\textbf{\textit{Data Structure Arrangements}}
 \vspace*{0.2cm}

For 2D simulations, data structure can be organized in memory in either a $(i,j,k)$ or a $(k,i,j)$ format \cite{DSL08}, with $(i,j)$ representing the (x,y) coordinates, and $k$ representing the direction. $(i,j,k)$ format optimizes the memory usage patterns in the streaming step, where all execution time is spent on data read/writes, whereas $(k,i,j)$ format optimizes for the memory access patterns in the collision step where the majority of execution time is spent waiting on the results of floating point calculations. This data structure arrangement provided up to a 40\% performance increase, and is part of the optimized serial performance shown in Table ~\ref{table:t8}. 

The fused streaming-collision method, common in LB optimization for single-phase flow simulations, reduces the required memory bandwidth for a node update in LB simulations, in which post-collision populations computed on the lattice grid layout are immediately streamed to the next neighboring nodes and stored on the second grid layout (i.e., on the second copy) and then these two grid layouts are swapped. Our code, however, does not incorporate the fused streaming-collision approach common, as our current implementation is not amenable to such an algorithmic change due to post-collision operations prior to streaming the populations (Eqs. ~\ref{e.MP7} $\mhyphen$ ~\ref{e.MP8} for Benchmark 1 and  Eqs. ~\ref{nn4} $\mhyphen$ ~\ref{nn5} for Benchmark 3), but future work may address this.

 \vspace*{0.2cm}
\noindent
\textbf{\textit{Task-level Pipelining}}
 \vspace*{0.2cm}

\begin{figure}[ht!]
\begin{center}
\includegraphics[width=12cm,height=12cm]{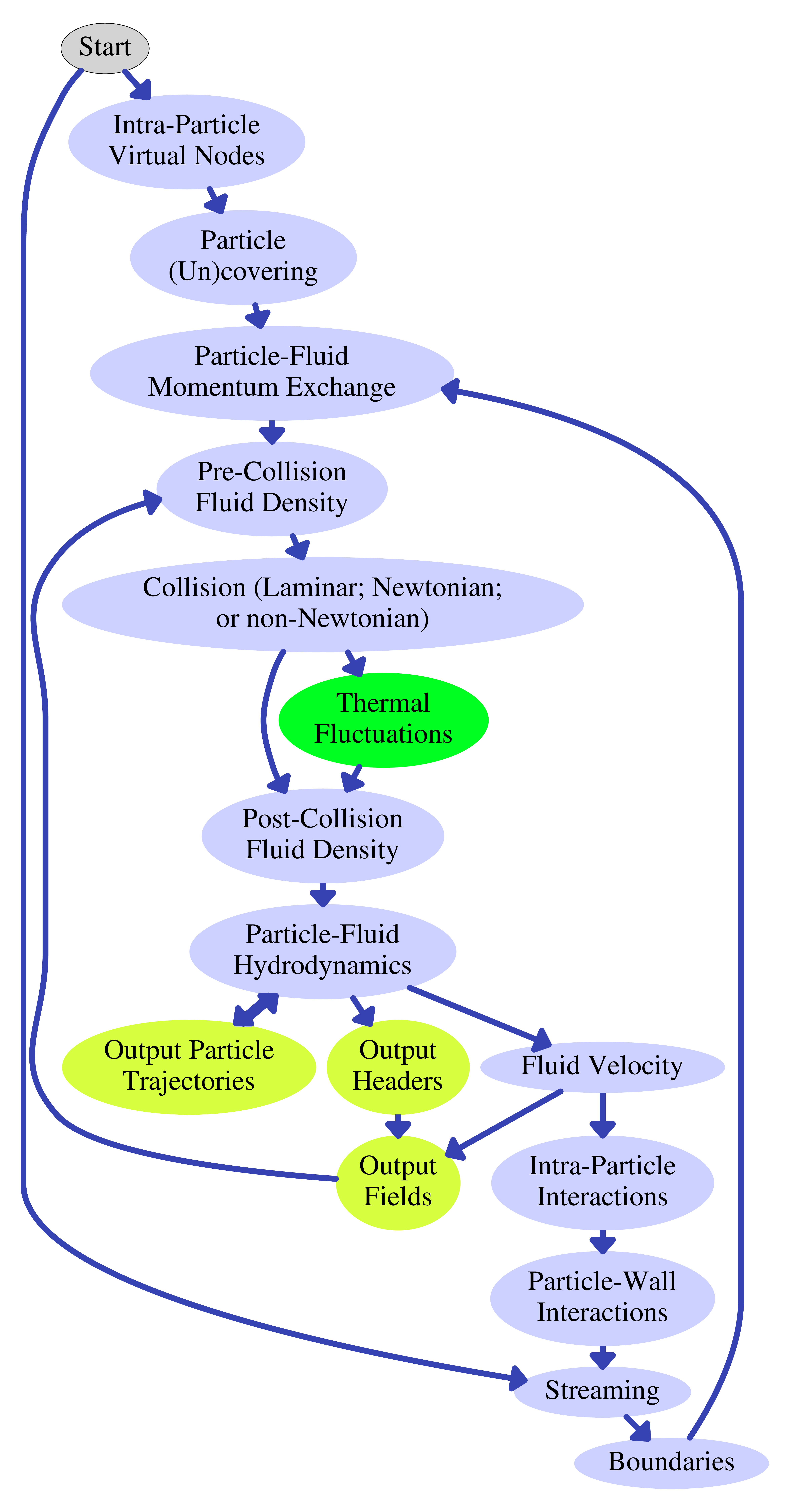}
\caption{Task level pipelining displaying the flow of data in the LB code. Modules shown in green were only used for some simulations. Modules shown in yellow handle simulation reporting.} \label{fig:pipeline}
\end{center}
\end{figure}

Building the rank ordering optimization of the main LB data structure, we added an additional dimension (time) to the population distribution rank (i.e $(i,j,k,t)$), similar to what is described in \cite{SSKDKSA13,MA02,PKOKNV08}. This eliminated the need for a $(t)$ and $(t+1)$ population distribution structure in our implementation, and enabled a further performance enhancement: efficient task-level parallelism/pipelining (Fig~\ref{fig:pipeline}), similar to the method in Ref. \cite{MSA14}, but within a single computational node. By codifying the data dependence graph below with Intel's Threading Building Blocks (TBB) library, we were able to achieve parallelism between many aspects of the LB model. For example, pipelining the collision$(t)$ and streaming$(t+1)$ actions increased performance 1.25$\times$ above the OpenMP code, and performing the particle interaction$(t)$ calculations with the streaming$(t+1)$ operation yielded additional increases of 10-20\% for simulations with large numbers of particles. In addition, by offloading simulation reporting from the OpenMP simulation threads (similar to the random number generate approach above), we were able to achieve an additional 2$\times$ increase beyond the OpenMP code for simulations with high reporting volumes. However, we also observed 10-20\% slowdowns on other simulations when we applied pipelining and/or offloading, demonstrating that the effectiveness of this technique is  dependent on simulation parameters. Therefore, fruitful runs utilizing this technique are not included in Table ~\ref{table:t8}.

\section{Summary and Conclusions}

A combination of different optimization techniques, when applicable, on the computational performance of the LB model for simulating three different problems were investigated. The problems studied involved (i) microparticle flows in a heterogeneous fracture granular-porous domain, (ii) fluid thermalization and the resultant Brownian particle motion important for submicron-scale particle flow simulations, and (iii) pseudoplastic fluid flow in a horizontal channel. In this paper, we applied these techniques to each benchmark problem separately, rather than to a problem involving all the modeling features simultaneously (e.g., simulating the flow of nano-size particle in a non-Newtonian fluid in a geometrically complex domain) to isolate the resultant speed-ups associated with each module. 

Applied techniques include optimization of array memory layouts, data structure simplification, random number generation outside the simulation thread(s), code parallelization via OpenMP, and intra- and inter-timestep task pipelining. As a result of implementation of these optimization techniques, we achieved significant speed-ups in the range of 14.58 (for the third benchmark problem) and 40.41 (for the second benchmark problem), with an average of $21 \times $ performance improvement in LB simulations. 

When Kuznik \textit{et al.} \cite{KOR10} ran their LB single phase flow simulation on 32 threads (closer to 28 threads were available to us) for a flow domain size of 256$\times$256, 512$\times$512, and 1,024$\times$1,024 lattice nodes (comparable to the number of lattice nodes in our Benchmark problems), they accomplished 86$\mhyphen$89 MLUPS. Because Kuznik \textit{et al.} simulated only a single-phase Newtonian fluid flow in a flow domain with no internal obstacles (i.e., there were no particles, thermal fluctuations, or non-Newtonian fluid flows), their reported speed-ups should be envisioned as an upper-bound for computational speed-ups on a single GPU. Our optimized LB code with computationally-involved thermal fluctuation calculations (Benchmark 2), however, performed $\sim$1.57$\times$ better than a single phase LB simulation on a single GPU. Similarly, our optimized code with non-Newtonian fluid flow simulations (computationally-demanding calculations for local velocity-dependent fluid kinematic viscosity in Benchmark 3) displayed nearly equal ($\sim$1.02$\times$ longer) computational performance with the single phase LB flow simulation on a single GPU. Moreover, when 45 particles were released into a flow domain that consists of internal obstacles, our optimized code resulted in a only $\sim$1.82$\times$ longer simulation, when compared to the LB simulation on a single GPU for a single-phase particle-free flow in a flow domain with no internal obstacles. Thus, the computationally-demanding LB simulations (advanced modeling features and/or complex flow domain geometries) in our benchmark problems on a CPU performed comparably or better than the single phase LB flow simulations on a single GPU reported in Ref. \cite{KOR10}. 

In regards to computational improvements through the use of GPUs in LB simulations of more complex flow problems, GPU/CPU speed-ups ranging from 2 to 12 were reported in LB simulations of multicomponent LB equations with multirange fluid-fluid interactions in flow domains of 128$\times$128 and 1,024$\times$1,024 lattice nodes \cite{BRBSS10}, which are comparable to the flow domain sizes in our benchmark problems, as reported in Table ~\ref{table:t9}. In addition, the computational performance of 53 MLUPS from a 3D LB simulation of particle-fluid dynamics in an irregular domain using a MPI-based parallelized code on C870 GPU \cite{BFM09} was at the lower end of the computational performances reported in Table ~\ref{table:t8}). Thus, our optimized CPU-based LB code performs comparably or better than previously reported multicomponent (fluid-fluid or fluid-particle) flow simulations on a single GPU. However, nearly an order of magnitude additional speed-up could be possible by changing the GPU cards, for example, moving from Fermi GPU to Pascal GPU without changing the code. Moreover, the authors in Ref. \cite{BFM09} reported 18x further speed-ups with the use of multiple GPUs as they move their simulations from C870 GPU to 8GT200 GPU, demonstrating the important role that state-of-the-art hardware platforms play in achieving maximum performance, and that cutting-edge GPUs can still outstrip cutting-edge CPU performance. This suggests that future work should investigate hybrid CPU-GPU systems in order to achieve maximum throughput and maintain scalability as problem size increases.

The aforementioned significant performance improvements (14.58$\mhyphen$40.41$\times$) could enhance the use of the LB method in simulating natural or targeted flows of nano-size and micron-size biotic (e.g., bacteria) or abiotic (engineered vectors or colloids) particles in geometrically complex flow domains. Applications may involve, for example, engineered targeted drug delivery to targeted tumor cells for cancer treatment, or health and safety assessments that utilize fate and transport of radionuclides bound to colloids in the vicinity of nuclear power plants or nuclear waste disposal sites.

In such cases, the computational performance and effectiveness of the aforementioned optimization techniques should be assessed on 3D simulations. The exponentially greater computational and memory requirements for such simulations means that the application of the techniques discussed in this paper is not straightforward, and will likely require modification in order to be successfully applied. For example, use of multi-processor system may become more feasible as memory requirements increase, necessitating revisions to our current OpenMP model. Furthermore, to target large 3D problems, multiple computational nodes will have to be utilized, requiring investigation into techniques of distributed memory sharing and task allocation on a inter-node level.

\label{sec:summary}

\section*{Acknowledgement}

 The work of H. Ba\c sa\u gao\u glu was supported by Internal Research Project 18-R8305. The  work of J. Harwell was supported in part by Internal Research Project 10-R8629. The authors thank Phil M. Westhart for his help with simulations and Miriam R. Juckett for reviewing the manuscript.


\end{document}